\documentclass{elsart}

\usepackage[authoryear]{natbib}
\usepackage{graphicx}

\newcommand{\gapprox}{$\stackrel {>}{_{\sim}}$}   
\newcommand{\lapprox}{$\stackrel {<}{_{\sim}}$}

\begin{document}
\begin{frontmatter}
\title{Interpreting the simultaneous variability of near-IR continuum and line emission in young stellar objects}

\author[mp]{D.Lorenzetti\corauthref{cor}},
\corauth[cor]{Corresponding author.}
\ead{dario.lorenzetti@oa-roma.inaf.it}
\author[mp]{S.Antoniucci},
\ead{simone.antoniucci@oa-roma.inaf.it}
\author[mp]{T.Giannini},
\ead{teresa.giannini@oa-roma.inaf.it}
\author[mp]{A.Di Paola},
\ead{andrea.dipaola@oa-roma.inaf.it}
\author[r1]{A.A.Arkharov},
\ead{arkadi@arharov.ru}
\author[r1,r2]{V.M.Larionov}
\ead{vlar2@yandex.ru}

\address[mp]{INAF - Osservatorio Astronomico di Roma - Via Frascati, 33 -
00040 Monte Porzio Catone, Italy}
\address[r1]{Central Astronomical Observatory of Pulkovo,
Pulkovskoe shosse 65, 196140 St.Petersburg, Russia}
\address[r2]{Astronomical Institute of St.Petersburg
University, Russia}

\begin{abstract}
 
We present new near-infrared (IR) spectra (0.80-1.35$\mu$m) of the pre-Main Sequence source PV Cep taken
during a monitoring program of eruptive variables we are conducting since some years. Simultaneous photometric and spectroscopic observations are systematically carried out during outburst and quiescence periods. By correlating
extinction-free parameters, such as HI recombination lines and underlying continuum, it is possible to infer
on the mechanism(s) responsible for their origin. Accretion and mass loss processes have a dominant role in
determining the PV Cep irregular variability of both continuum and line emission. The potentialities of the
observational modality are also discussed.

\end{abstract}


\end{frontmatter}

\section{Introduction}

Young stellar objects (YSO's) present a certain degree of irregular variability which is typically 
attributed to intermittent accretion (or mass loss) phenomena. According to a widely accepted picture 
(Hartmann \& Kenyon, 1985; Shu et al. 1994), accretion matter viscously migrates through the 
circumstellar disk toward its inner edge from where this matter is intermittently channeled along 
the magnetic field lines to the central star. The fall onto the stellar surface
produces a shock that cools by emitting a hot continuum; moreover, as a consequence of the 
accretion event, strong winds (in some cases also collimated jets) emerge from the rotating 
star/disk system. Accretion-related variability may be due also to dust evaporation and condensation, a 
phenomenon potentially relevant in V1647 Ori (Reipurth \& Aspin 2004, Acosta-Pulido et al. 2007, Aspin 2011); 
V2492 Cyg (Hillenbrand et al. 2012); and PV Cep itself (Kun et al. 2011). 
Irregular variability is alternatively attributed to a variable extinction caused 
by dust clumps that occasionally intersect our line of sight (Skrutskie et al. 1996). Both mechanisms 
are plausible because all YSO's are expected to accrete, although with decreasing rates while the 
evolution goes on, and they are preferentially located in the densest parts of molecular clouds where gas 
and/or dust clumps are usual ingredients.\ 

The classical observational method for studying the modalities and the causes of the irregular variability 
consists of monitoring the continuum emission in different spectral bands to obtain the 
light-curves and the colors of the fluctuations. A further (and even better) technique is 
to exploit simultaneously a photometric and spectroscopic monitoring in the same spectral band: 
this represents a powerful tool of investigation since provides extinction-free parameters that allow us 
to understand {\it (i)} the mechanism(s) responsible for the continuum and line emission; {\it (ii)} 
whether or not these mechanisms are (inter-)related; and {\it (iii)} in affirmative case, to infer on the 
possible common mechanism.\ 

Remarkably, since few years we are conducting a photometric (Lorenzetti et al. 2006, 2007) and spectroscopic 
(Lorenzetti et al. 2009) monitoring in the near-IR (0.8-2.5 $\mu$m, ZJHK bands) of a sample of low-mass pre-main sequence 
sources named EXor's (Herbig 1989; Reipurth \& Aspin 2010).  These are targets very suitable for the aim described above, 
since they present large (about 3-4 mag) outbursts, characterized by a short (weeks, months) rising and a longer (months, 
years) declining, superposed to long (years) quiescence periods. In particular, the source PV Cep was extensively monitored (Lorenzetti et al. 2011; see also for a detailed description of this pre-Main Sequence object); its variability, firstly 
interpreted as due to extinction, has been later reconsidered as due to accretion and our aim here is
to demonstrate (in the specific case of PV Cep) the effectiveness of the proposed approach. The novelty of the present 
short note relies on both presenting new photometric and spectroscopic data and, most importantly, considering new 
diagnostic parameters. Obviously, the approach adopted here for a single object (PV Cep) and for a relatively short 
monitoring period (about 3 years) can be ported in a more general context, once provided a meaningful 
monitoring of other sources.\ 

This paper is organized in the following way: Sect.2 shortly refers to the used instrumentation; the relevant 
parameters derived from previous and new observations are summarized in Sect.3; our considerations are presented 
in Sect.4, while final remarks are given in Sect.5.

\section{Near-IR photometry and spectroscopy}

The following analysis is based on near-IR photometric and spectroscopic data. These latter
were obtained at the 1.1m AZT-24 telescope located at
Campo Imperatore (L'Aquila - Italy) equipped with the
imager/spectrometer SWIRCAM (D'Alessio et al. 2000), which is
based on a 256$\times$256 HgCdTe PICNIC array. Photometry is
performed with broad band filters J (1.25 $\mu$m), H (1.65
$\mu$m), and K (2.20 $\mu$m). The total field of view is
4.4$\times$4.4 arcmin$^2$, which corresponds to a plate scale of
1.04 arcsec/pixel. All the observations were obtained by dithering
the telescope around the pointed position. The raw imaging data
were reduced by using standard procedures for bad pixel removal,
flat fielding, and sky subtraction.

Low resolution ($\mathcal{R}$ $\sim$ 250) spectroscopy is obtained
by means of two IR grisms G$_{blue}$ and G$_{red}$ covering the ZJ
(0.83 - 1.34 $\mu$m) and HK (1.44 - 2.35 $\mu$m) bands,
respectively, in two subsequent exposures. The long slit is not
orientable in position angle, and it samples a pre-defined portion
of the focal plane, 2$\times$260 arcsec$^2$ in the east-west
direction.

Long-slit spectroscopy was carried out in the standard
ABBA' mode with a total integration time ranging
between 800 and 1200 sec. The spectral images were flat-fielded,
sky-subtracted, and corrected for the optical distortion in both
the spatial and spectral directions. Telluric features were
removed by dividing the extracted spectra by that of a normalized
telluric standard star, once corrected for its intrinsic spectral
features. Wavelength calibration was derived from the OH lines
present in the raw spectral images, while flux calibration was
obtained from our photometric data carried out on the same night
of the spectroscopy.\\

\begin{figure}
\includegraphics[width=14cm]{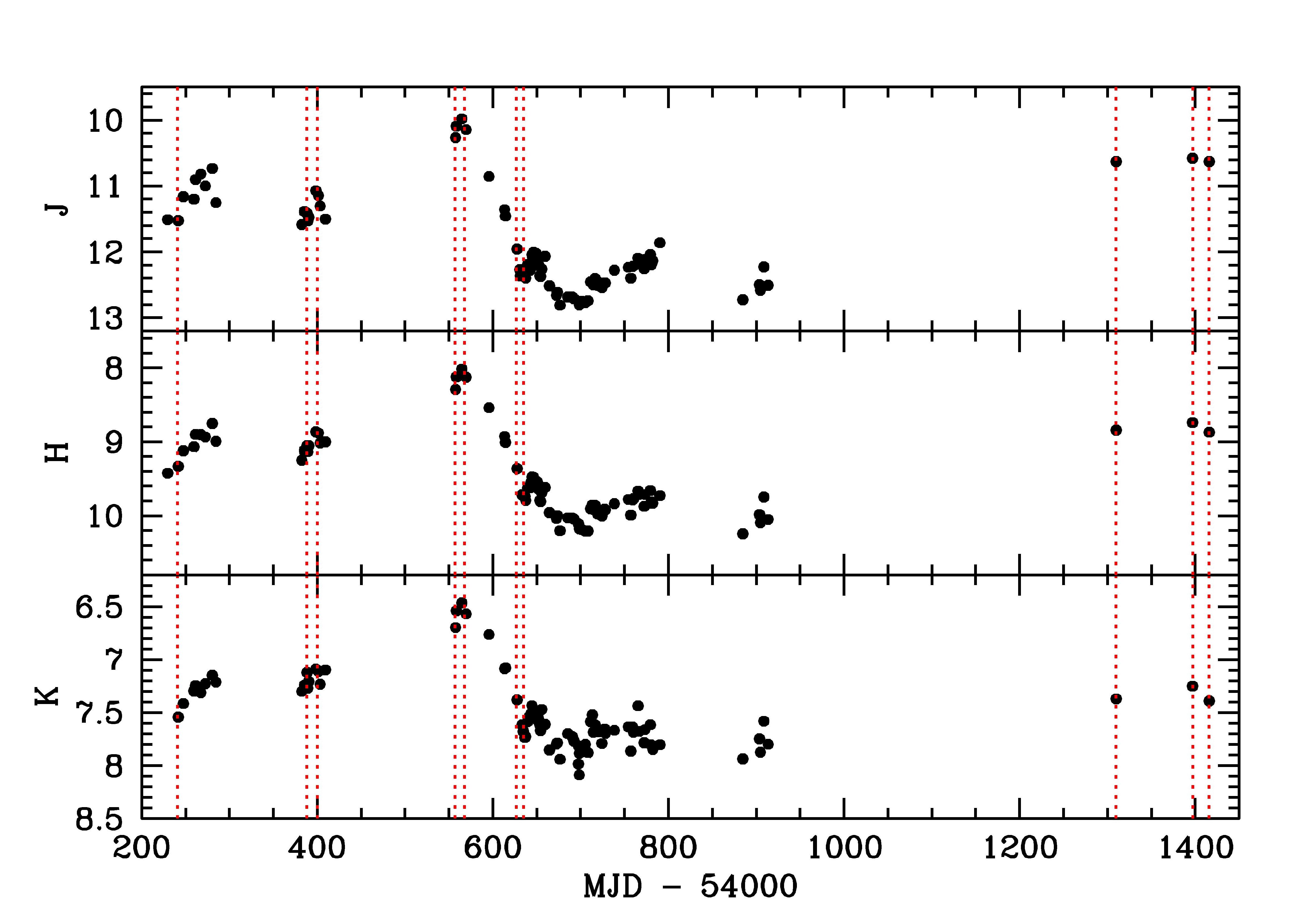}
\caption{\label{lightcurve:fig} PV Cep near-IR light curves vs. MJD (Modified Julian Date).
  The errors of data points are comparable to or lesser than 0.03 mag. Vertical red lines
  correspond to those dates when, beside JHK photometry, also a near-IR spectrum was taken.}
\end{figure}

During the monitoring period presented here, the near-IR
spectra of PV Cep were taken in several occasions marked in
Figure~\ref{lightcurve:fig} with red vertical dotted lines.
Since this object is a target included in a photometric and
spectroscopic monitoring program we are performing since few years, 
some of these spectra have been already discussed elsewhere (Lorenzetti et al. 2009). The
oservational novelty is represented by the 3 new spectra (and relative photometry) 
obtained in April 2010 (MJD 55310), July 2010 (MJD 55397), and
August 2010 (MJD 55416) (see Figure~\ref{lightcurve:fig}). They will be commented,  
along with previous data, in the following Sect.4.

\section{Results}

\begin{figure}
\includegraphics[width=14cm]{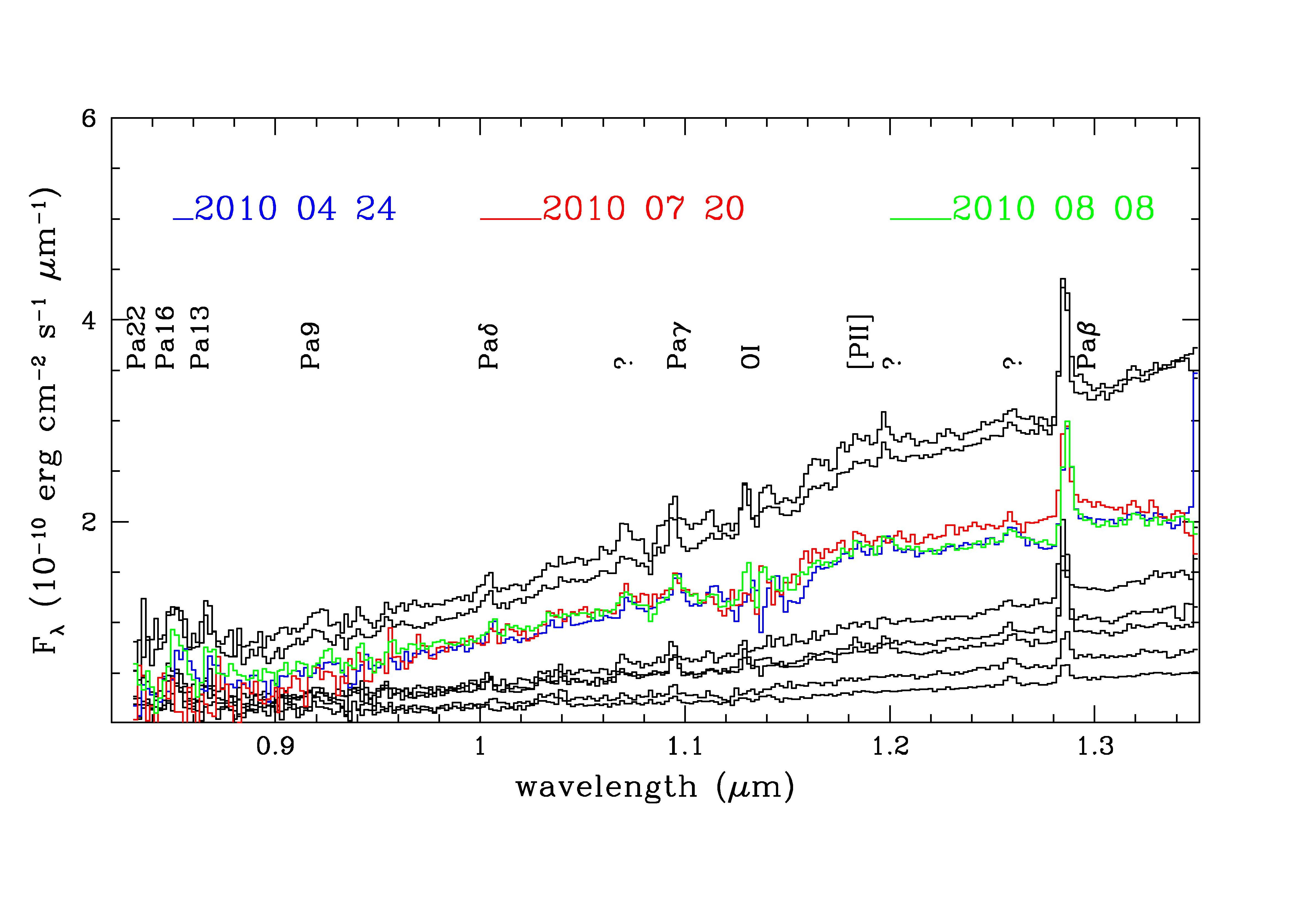}
\caption{\label{spectra:fig} The newly obtained near-IR spectra of PV Cep are depicted in
   colours (red, green, magenta). For comparison purposes, the already presented spectra 
   are also plotted (in black).}
\end{figure}

The ZJ portion of the newly obtained (during 2010) near-IR spectra of PV Cep is
plotted in Figure~\ref{spectra:fig} along with our previous results.
New data substantially confirm the existence of some continuum variability. HI recombination 
lines of the Paschen series are present: among them the Pa$\beta$
appears as the most prominent in all spectra and on that line we will focus our attention.
In Table~\ref{data:tab}, the Pa$\beta$ observed flux, its equivalent width (EW) and the value
of the underling J band continuum (in mag) are given for each date. Given the spectral closeness between
the J-band effective wavelength (1.25 $\mu$m) and  the Pa$\beta$ transition (1.28 $\mu$m),
the EW of this latter can be considered unaffected by the extinction. As such, EW is a suitable
parameter to give indication on whether or not, and to what extent, the mechanisms responsible for
the continuum and line emission are related (see next Sect.). The same is obviously true also for other 
pairs of HI recombination lines/underlying continuum: indeed we did the same analysis also 
for Br$\gamma$/K band combination obtaining the same results not reported here to avoid any 
unuseful redundance.

\begin{table}
\begin{center}

\caption{Spectroscopic and photometric parameters of PV Cep. \label{data:tab}}
\medskip
{\normalsize
\begin{tabular}{ccccc}
\hline
\medskip
Date         &    MJD$^a$   &  Pa$\beta$ Flux                    &   EW          &    J band  \\
             &              & (10$^{-13}$ergs$^{-1}$cm$^{-2})$   &  (\AA)        &     (mag)  \\ 
\hline
\medskip 
May 13, 2007 &    54241     &   3.8 $\pm$ 0.2                    & -48 $\pm$ 2.4 &  11.52$^b$ \\
Oct 15, 2007 &    54389     &   3.7 $\pm$ 0.2                    & -40 $\pm$ 2.0 &  11.53     \\ 
Oct 27, 2007 &    54401     &   3.5 $\pm$ 0.2                    & -28 $\pm$ 1.7 &  11.14     \\
Apr 01, 2008 &    54558     &   5.3 $\pm$ 0.2                    & -16 $\pm$ 0.6 &  10.09     \\ 
Apr 12, 2008 &    54569     &   6.1 $\pm$ 0.2                    & -20 $\pm$ 0.6 &  10.14     \\ 
Jun 10, 2008 &    54627     &   1.4 $\pm$ 0.2                    & -22 $\pm$ 3.0 &  11.96     \\
Jun 18, 2008 &    54635     &   1.2 $\pm$ 0.2                    & -19 $\pm$ 3.0 &  12.34     \\
Apr 24, 2010 &    55310     &   5.5 $\pm$ 0.2                    & -29 $\pm$ 1.3 &  10.63     \\
Jul 20, 2010 &    55397     &   5.1 $\pm$ 0.2                    & -25 $\pm$ 1.6 &  10.58     \\ 
Aug 08, 2010 &    55416     &   5.9 $\pm$ 0.2                    & -31 $\pm$ 1.8 &  10.63     \\        
\hline  

\end{tabular}}
\end{center}
\medskip
$^a$ MJD = Modified Julian Day\\
$^b$ Errors on J magnitudes are typically 0.02-0.03 mag

\bigskip

\end{table}

\section{Analysis and discussion}

\begin{figure}
\includegraphics[width=14cm]{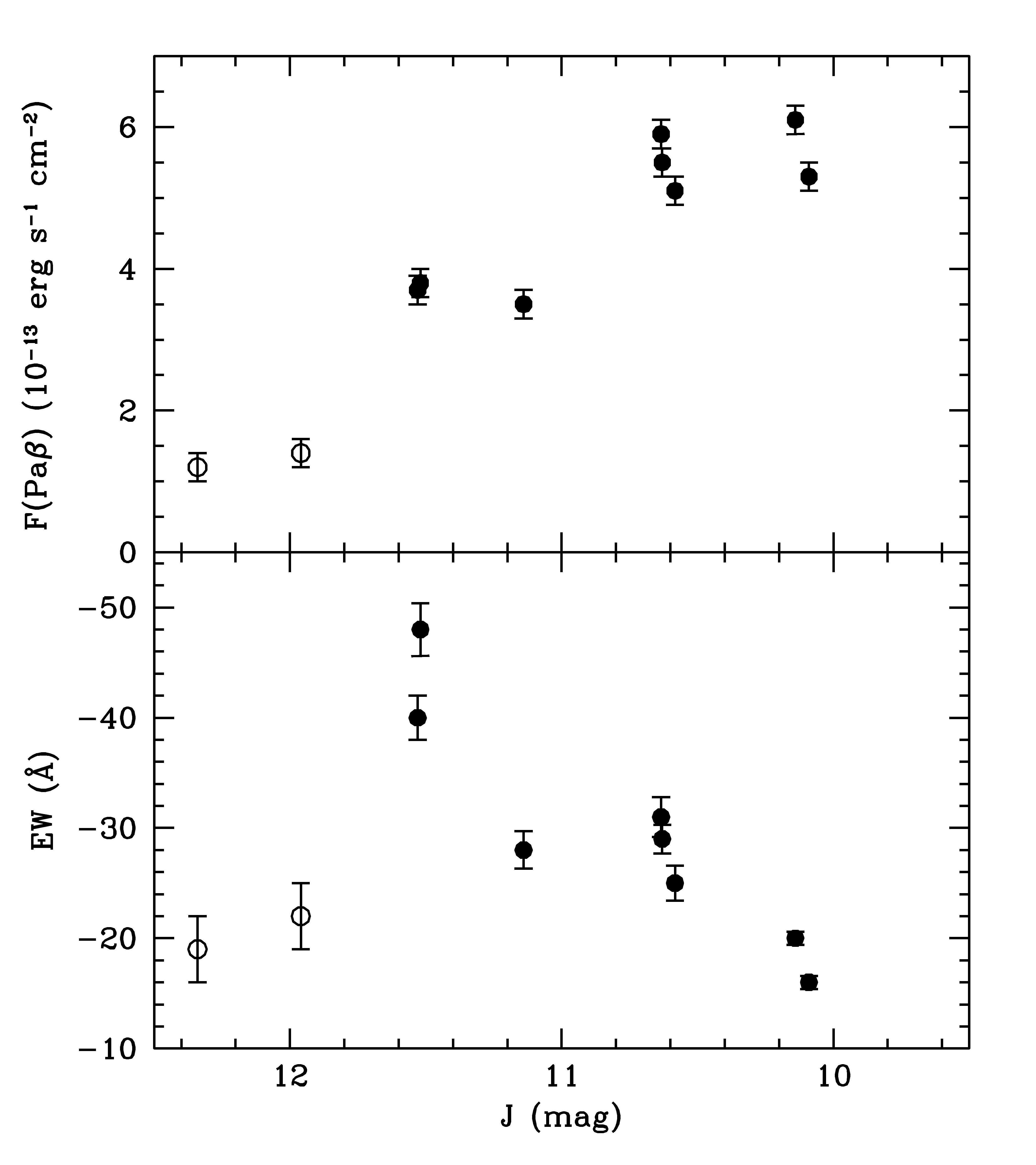}
\caption{\label{EW_J:fig} {\it Upper panel.} Distribution of the Pa$\beta$ fluxes extracted from the different
   spectra of PV Cep as a function of its J magnitude.
   {\it Lower panel.} Same as before, but in terms of Equivalent Widths;
   as usual, EW have negative values because line are in emission.
   Open symbols refer to the spectra taken just after a rapid
   fading of the source (see text and Figure~\ref{EW_time:fig}).}
\end{figure}

In the upper panel of Figure~\ref{EW_J:fig} an evident correlation (regression coefficient = 0.93)
is recognizable between the Pa$\beta$ line flux and the continuum
in the J band. Such a correlation suggests, as expected, that emission lines are
generated by some mechanism strongly related to the central source.
According to the lower panel of the same figure an
anti-correlation seems to exist between EW and J continuum, in the 
sense that decreasing EW values correspond to an
increasing continuum brightness. Two data points, those
corresponding to J about 12 mag and marked with open symbols,
seem not to follow the anti-correlation trend, but this apparent 
contraddiction will be discussed later. Neglecting these two points, 
a regression coefficient of 0.91 is obtained. Now we remark 
the anti-correlation indicates that, although line emission and J continuum
are related, the latter presents an increase larger (i.e. faster) than the
former. Indeed, the J-band flux is composed of the {\it direct} radiation from the hot spot and an
{\it indirect} brightening due to the radiation from the innermost regions of the circumstellar disk. Since 
PV Cep is embedded in an extremely massive disk (Hamidouche 2010), seen at large inclination, this effect has to be 
accounted for and likely represents the additional contribution that makes the continuum variation larger than
the line emission. Similar anticorrelation results are given by Cohen et al. (1981), Magakian \& Movsessian (2001), and Acosta-Pulido et al. (2007).\
 
\begin{figure}
\includegraphics[width=14cm]{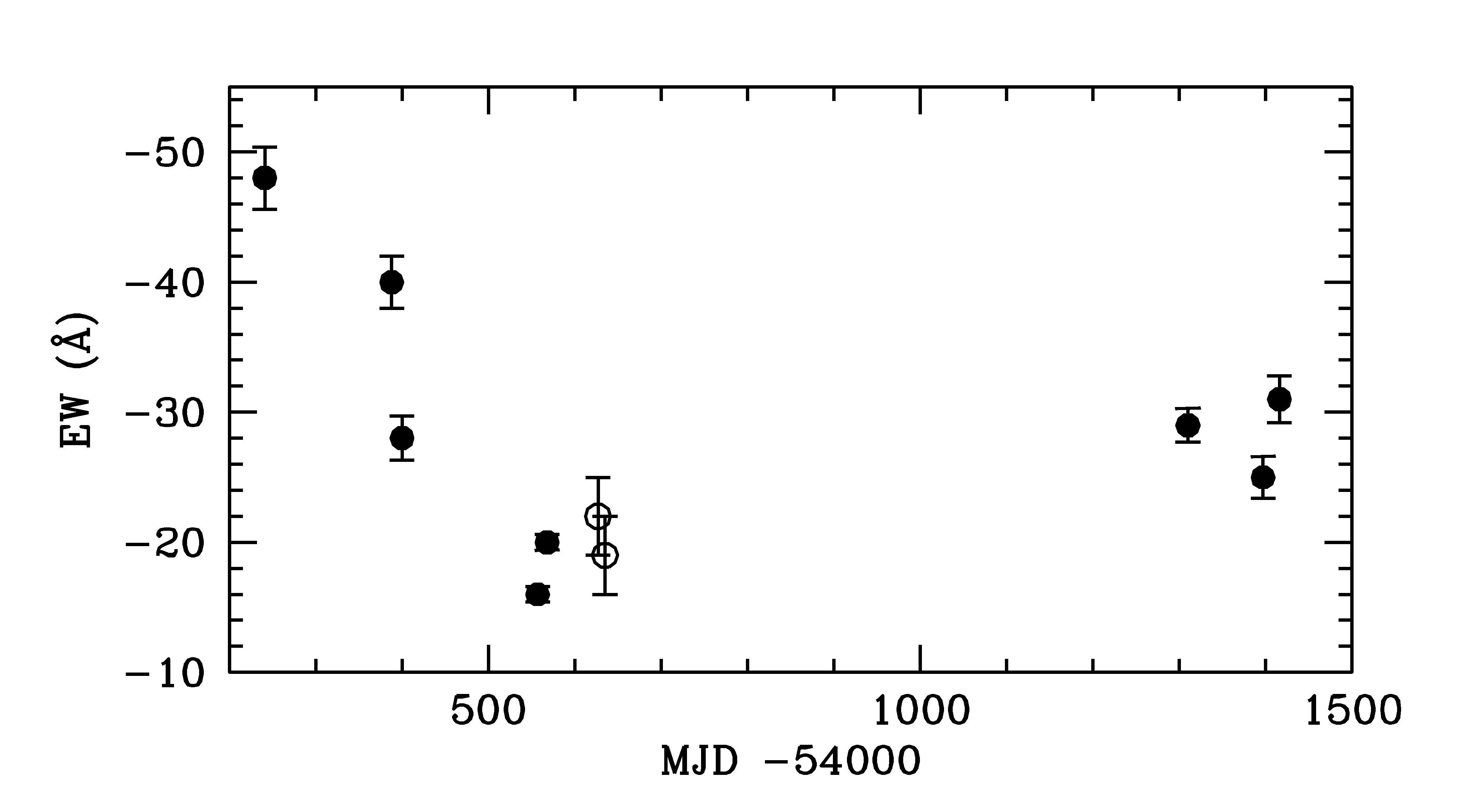}
\caption{\label{EW_time:fig} EW derived from Pa$\beta$ spectroscopy as a function of time. As in Figure~\ref{EW_J:fig} the same couple of data points is represented with open symbols.}
\end{figure}

This occurrence tends to rule out the variable extinction as
cause of the observed variability that affects to the same extent both lines and continuum, since in that case a constant value
of the EW should be expected for any continuum fluctuation. Even more a selective obscuration can be also ruled out. In fact, it should be located
at the dust condensation zone, very close to the star and, as such, it should obscurate the stellar photosphere more than the accretion columns
and the wind regions. On the contrary,
accretion (or mass loss) processes are compatible with the observed behaviour, since they 
respond with different laws (Muzerolle et al. 1998), and maybe with different lags, to the continuum 
variations. These latter, once ruled out any extinction variation are unanimously attributed to
hot spots on the stellar surface due to the shocks that originate at the base to the 
columns of the infalling matter. Being the line fluxes well correlated with the increasing 
continuum, they will depend on accretion processes: such a dependence may be direct if 
lines originate in the accretion columns, or indirect if lines originate in a stellar wind
whose mass loss rate ($\dot{M}_{wind}$) is related to the mass accretion rate ($\dot{M}_{acc}$)
by invoking the rough proportion $\dot{M}_{wind}$/$\dot{M}_{acc}$ \lapprox~ 0.1 (Shu et
al. 2000; K\"{o}nigl \& Pudritz 2000). Therefore we can assume that the total HI recombination line flux
is the sum of different contributions: from accretion, from wind and, possibly, from other components 
(e.g. an HII circumstellar region). During strong accretion events, that increase the source continuum emission,
only one contribution to the line emission dominates (correlation depicted in Figure~\ref{EW_J:fig}), but when
the impulsive accretion stops, namely when the continuum drops below a given threshold 
(in the present case J mag \gapprox~ 12), line emission strength comes back to its original (lower) value,
practically independent on the continuum value, and the EW value suddenly changes (see open symbols in Figure~\ref{EW_J:fig}). 
In principle, alternative interpretations could exist: the proposed one is well compatible with the observational data and 
reconciles the two apparently unconsistent data points. However, to firmly confirm the presented scenario we still needs additional data resulting from the photometric and spectroscopic monitoring of relatively strong (more then 1-2 mag) variations of PV Cep or other targets. Indeed, one aspect that would deserve a deeper investigation is the following: by looking at Figures~\ref{EW_time:fig} and \ref{lightcurve:fig}, we can realize that the EW value reached immediately after the accretion event (that indicated by open symbols) is very similar to that corresponing to observations taken 100 days before, at an outburst peak. According to the proposed scenario, any EW value is associable
to a continuum value lesser than a threshold, hence the observed EW resemblance is remarkable, but, plausibly, fortuitous.

 \section{Final remarks}
 
The case presented here for a single young source (PV Cep) aims at fostering systematic photometric 
and spectroscopic monitoring intended as an observational method of investigation on how the 
mechanisms responsible for continuum and line emission are (inter-)related. Unfortunately, the
proposed technique is not widely exploited since a frequent (weekly) monitoring (on years time-scale) is 
not easily doable both on large telescopes affected by a heavy oversubscription and on smaller 
telescopes often not equipped with the needed instrumentation. It would be very fruitful if
small telescopes (1-2m class), but properly equipped, were dedicated to similar projects even for
a small fraction ($<$ 10\%) of their observing time.

\vspace{3cm}

\end{document}